\documentclass{aa} 
\usepackage[dvips]{graphicx}
\usepackage{txfonts}
\usepackage{natbib}
\bibpunct{(}{)}{;}{a}{}{,}

\newcommand{\gt}{>}
\begin{document}
\title{Kinematics of Vega-like stars: lifetimes and temporal evolution
of circumstellar dust disks}
\author{Manoj P. \and H. C. Bhatt}
\offprints{Manoj P.}
\institute{Indian Institute of Astrophysics, II Block, Koramangala, Bangalore - 560034\\
            \email{manoj@iiap.res.in}}

\abstract{ We have used the velocity dispersion as an age indicator to
constrain the ages of a large sample of main sequence Vega-like stars
in order to study the lifetimes and temporal evolution of the dust
disks around them. We use {\it Hipparcos} measurements to compute
stellar velocities  in the sky plane. The velocity dispersion of
Vega-like stars is found to be systematically smaller than that of the
normal main sequence field stars for all spectral types, suggesting
that the main sequence dusty systems, on  average, are younger than
 normal field stars.  The debris disks seem to survive longer
around late type stars as compared to early type stars. Further, we
find a strong correlation between fractional dust luminosity (\(
f_{d}\: \equiv \: L_{dust}/L_{\star } \)) and velocity dispersion of
stars with dust disks. Fractional dust luminosity is found to drop off
steadily with increasing stellar age from early pre-main sequence
phase to late main sequence phase.

} 

\titlerunning{Lifetimes and evolution of main sequence dust disks}
\authorrunning{Manoj P. \& H. C. Bhatt}
\maketitle

\section{Introduction}

It has been well established that the  majority of pre-main sequence
stars are surrounded by circumstellar disks which are analogous in
their properties to the protosolar nebula before the onset of planet
formation \citep[e.g.,][]{beck99, mlw00, wl00,becksar96}. Young
circumstellar disks lose most of this material due to planet formation
and other disk dispersal processes by the time the central stars
harbouring these disks reach the main sequence. In the standard model
of planet formation, the dust grains with sizes typical of
interstellar dust settle down to the disk mid-plane and stick
together to grow into rocky planetesimals \citep[e.g.][]{wc93}. The
disk is depleted of smaller grains and this lowers the opacity of the
reprocessing disk. Planetesimals grow further to earth-like planetary
bodies by coalescence and eventually accrete gas in the outer disk to
form giant planets \citep[e.g.][]{pollack96}. When the disk has
become sufficiently gas-free, so that it is dominated by grain
dynamics, the planetary mass objects can gravitationally perturb
kilometer-sized planetesimals sending them into highly eccentric
orbits. Collisions between these planetesimals then replenish the
disk with `second generation dust' which is observed around many main
sequence stars \cite[e.g.][]{lba00}. These main sequence stars with
debris disks, known as `Vega-like' stars, were first discovered by
IRAS in 1983 \citep{aum84}. At least 15\% of  main sequence stars
are surrounded by such disks \citep{lba00}.

Planet formation is well underway in main sequence dusty systems and
the disks that we observe are the debris product of the planet
formation process \citep{lba00}. These disks are thought to be `sign
posts' of planet formation \citep{kenbro02}. What are the lifetimes of
these disks? Do disk lifetimes depend on the nature of the central
star? How do these observed disks evolve with time?  These questions are
central to our understanding of planet formation and disk evolution. A
study of the lifetimes and the temporal evolution of the dust disks
should provide insight into the formation of planetary systems and
disk dispersal timescales and mechanisms.

There have been a number of studies on the evolution of circumstellar
disks around main sequence stars. \citet{zb93} have found that the
mass of dust in the disks declines as rapidly as \( (time)^{-2} \)
during the initial \( 3\times10 ^{8} \) yr. Similar results have been
reported by \citet{hol98} from their SCUBA observations.  There
is general agreement now on the fact that the amount of dust retained
in the disks decreases with increasing stellar age \citep{hb99,lba00}.
However, the exact nature of this decline is not clear. Most of these
studies are based on the ages of a few prototype Vega-like stars.
When more stars are employed, reliable estimation of their ages poses
a serious problem. 

It is difficult to determine the ages of main sequence stars with
reasonable accuracy. There have been a few attempts to estimate ages
of field Vega-like stars \citep[e.g.][]{lauch99,song00,silv00}, but the
ages determined using different techniques are not always mutually
consistent \citep{zuck01}. Recently, \citet{spang01} have carried out
a survey of circumstellar disks around pre-main sequence and young
main sequence stars that are members of young open clusters of known
ages using ISOPHOT. They found that the fractional dust luminosity \(
f_{d} \) drops off with stellar age according to the power law \(
f_{d}\propto (age)^{-1.76} \).  This suggestion of a global power law
has been contested by \citet{decin03} who find a spread in fractional
dust luminosity at any age from the revised age estimates of their
sample stars observed by ISOPHOT.  \citet{dom03}, based on a physical
model that they developed for the dust production in Vega-like disks,
have argued that a collisional cascade with constant collision
velocities leads to a power law decrease of the amount of dust seen in
the debris disk with a power law index of $-1$. They add that a collisional
cascade with continuous stirring can produce slopes steeper than $-1$.

 In this paper, we study the temporal evolution of dust disks around
 main sequence stars. We consider the kinematics of a large sample of
 Vega-like stars and use the velocity dispersion as an age
 indicator. It has long been known that there is a strong correlation
 between the random velocities and ages of stars in the Galactic
 disk. Velocity dispersion of stars in the solar neighborhood has been
 found to increase with  age \citep{wiel77, jw83}. Observationally,
 velocity dispersion \( \sigma \) is found to grow with age at least
 as fast as \( t^{0.3} \) and more likely \( t^{0.5} \) \citep{wiel77,
 bt87}. The dynamical origin of this effect is attributed to the
 encounters between the disk stars and the massive gas-clouds
 \citep{spschw51,spschw53} and to transient spiral waves heating up
 the Galactic disk \citep{bw67}. Using accurate \textit{Hipparcos}
 parallaxes and proper motions, \citet{bdb00} and \citet{db98} have
 shown that for a coeval group of stars, the {\it rms} dispersion
 in transverse velocity S (in the plane of the sky), which is
 connected to the principal velocity dispersion by the relation \(
 S^{2}=2/3[\sigma _{R}^{2}+\sigma _{\phi }^{2}+\sigma _{z}^{2}] \),
 increases with time from 8 kms$^{-1}$ at birth as \( t^{1/3} \).  We
 follow this formalism and use the dispersion in transverse velocity
 to constrain the ages of Vega-like stars in order to study the
 lifetimes and temporal evolution of the dust disks.

\section{Data}

A number of recent studies give lists of Vega-like stars and candidate
Vega-like stars, selected on the basis of their infrared excesses in
the \textit{IRAS} wave bands. \citet{song00} lists 361 objects taken
from different surveys and searches published in the literature.  From
a search of the \textit{IRAS} FSC catalog, \citet{silv00} produced a
list of 191 Vega-like stars. A number of additional Vega-like objects
have been discussed in \citet{coul98} and \citet{malf98}. We first
compiled a total of 486 distinct Vega-like stars taken from these
lists that had many objects in common. This large sample could have
some stars that are erroneously classified as Vega-like or have
uncertain associations with the \textit{IRAS} sources due to the large
\textit{IRAS} beam size. In our study, we consider only those stars
for which the positional offset between the optical star and the
\textit{IRAS} association is \( \leq 30^{\arcsec } \). Further,
we examined the Digital Sky Survey images of the region near each
Vega-like candidate to make sure that the  far-infrared emission is indeed
from the stellar source and not from any extended background source,
for example a galaxy, near the star in the plane of the sky
\citep[e.g.,][]{zs04}. We have eliminated those stars from our sample
where an optical galaxy or an \textit{IRAS} extended source was found
inside or very near ( $\leq$ 1 $\arcmin$ ) to the \textit{IRAS} error
ellipse for the point source.  Alternative associations have been found
for some of the proposed Vega-like stars in the literature
\citep{sylv00, liss02}.  We have excluded such stars from our
sample. We also exclude known pre-main sequence stars (e.g., Herbig
Ae/Be stars; \citet{the94}) and other emission-line objects from our
sample. Further, we consider only stars in the spectral range between
B9 and K5 - infrared excess from early B type star could be due to
free-free emission \citep{zuck01} and K-giants are known to exhibit
Vega-like excesses \citep[e.g.,][]{plets97, jura99}. Finally, our
sample contains 221 Vega-like stars for which both \textit{Hipparcos}
and \textit{IRAS} (PSC/FSC) measurements are available.

\section{Analysis}

\subsection{Dustiness of Vega-like stars}

A good measure of the `dustiness' of the disks around Vega-like stars
is the fractional dust luminosity \( f_{d}\; \equiv \;
L_{dust}/L_{\star } \), which represents the optical depth offered by
an orbiting dust disk to ultraviolet and visual radiation
\citep{zuck01}. We compute \( f_{d} \) from \textsc{IRAS (PSC/FSC)}
fluxes for the Vega-like stars in our sample using the relation \[
f_{d}=L_{dust}/L_{\star }=\frac{10^{-4}\times
[6.45e_{12}+2.35e_{25}+1.43e_{60}+0.55e_{100}]}{10^{[0.4(4.75-m_{V}-BC)]}}\]
\citep{emers88}. In the above equation \( e_{12},\; e_{25},\;
e_{60},\; e_{100} \) are the excess flux densities in Jy over the
photospheric values in the IRAS wave bands at 12, 25, 60 and 100\( \mu
\)m respectively, \( m_{V} \) is the visual magnitude of the star
corrected for extinction (which is generally very small, typically
$\le$ 0.1 mag for stars in our sample) and BC  is  the bolometric
correction.

The excess flux density in each {\it IRAS} band was estimated as
follows. The photospheric 12$\mu m$ magnitude was derived from the
extinction-corrected V magnitude and (B-V) color of the star as
discussed in \citet{oudmaijer92}. The photospheric magnitudes at the
other bands were calculated using the relations given in
\citet{oudmaijer92}. Photospheric magnitudes were then converted into
flux densities in the {\it IRAS} bands by using the magnitude zero
points listed in the {\it IRAS} catalog, and then color corrected to a
photospheric SED \citep[e.g.,][]{silv00}. The photospheric estimates
in the {\it IRAS} catalog color convention were subtracted from the
corresponding non-upper limit {\it IRAS} PSC/FSC flux densities to
obtain the excess flux densities $e_{\lambda}$ in each of the {\it
IRAS} bands.

In order to account for the possible inexact approximation of the
photosphere, the excesses computed are considered, as in
\citet{silv00}, to be significant only if it exceeds 20\% of the
photospheric flux value in each of the bands. Stars with significant
excess in any one of the bands are taken as `true' Vega-like and their
$f_d$ is computed as described above. Fractional dust luminosities
thus computed for 181 `true' Vega-like systems agree well with earlier
estimates in the literature \citep{backgill87,song00, silv00},
generally to within 10\%.  We find a number of stars with excesses
of less than 20\% of the photospheric fluxes in all the four {\it IRAS}
bands. The excesses, if any, that these stars show are at a very low
level. However, these stars have been classified as Vega-like stars in
earlier studies in which excesses were inferred by different
methods. We assign an upper limit value of $10^{-6}$ for $f_d$ of
these stars as the values of $f_d$ computed for them are
$\la\:10^{-6}$.

\subsection{Kinematics - transverse velocities of Vega-like stars}

For all the stars in our sample we have proper motions and parallaxes
from the \textsl{Hipparcos} catalog \citep{esa97}. The transverse velocity
perpendicular to the line of sight relative to the solar system
barycenter is then computed using the relation \( V_{T}\; =\;
\frac{A_{v}\: \mu }{\pi } \) where $A_{v}\: =\: 4.740470$ km yr
s$^{-1}$, \( \pi \) is the parallax in {\it milliarcseconds} and \(
\mu =({\mu _{\delta }^{2}\: +\: (\mu _{\alpha }\cos \delta
)^{2}})^{1/2} \) with \( \mu _{\delta } \) and \( \mu _{\alpha }\cos
\delta \) being the proper motions along declination and right
ascension in {\it milliarcseconds}.  Errors in transverse velocities
are estimated from the probable errors in parallaxes and proper
motions given in the \textit{Hipparcos} catalog. Transverse velocities
of stars thus obtained will have solar motion reflected in them. We
have corrected the velocities for solar motion using the values of
$U=10.0\pm 0.4$ km~ s$^{-1}$, $V=5.2\pm 0.6$ km~s$^{-1}$, $W=7.2\pm
0.4$ km~s$^{-1}$ \citep{bm98} for the standard solar motion.  To
minimize the effect of Galactic differential rotation we consider only
stars within 250~pc from the Sun. Further, we include only those stars
in our analysis that have a fractional error in transverse velocity less
than \( 0.5 \). We have, then, \( 158 \) Vega-like stars with
transverse velocities and fractional dust luminosities computed for
the final analysis.

\section{Results}

\subsection{Velocity dispersion and disk lifetimes}

As discussed in $\S$1, the velocity dispersion of a group of stars is
a measure of the average age of the group. Here we use transverse
velocity dispersion as an age indicator to constrain the ages of the
stars with dusty disks in order to study the disk lifetimes and
evolution. The  transverse velocity dispersion of the 158
Vega-like stars in our final sample is found to be 21.4 $\pm$ 1.2
km~sec$^{-1}$. The transverse velocity dispersion, again
computed using {\it Hipparcos} measurements, of about 14 000 field
stars  with spectral types between B9 and K5 which are within 250~pc
from the Sun and whose fractional errors in transverse velocities are
less than 0.5, is computed to be 37.3 $\pm$ 0.2 km~sec$^{-1}$. The
smaller velocity dispersion for Vega-like stars compared to field
stars indicates, at the very outset, that the main sequence dusty
systems are younger than the field stars and that the debris disk
lifetimes are shorter than the main sequence lifetimes of the
stars. In the following, we analyze the spectral type dependence of
the velocity dispersion of Vega-like stars.

\begin{figure}
\centering
\resizebox{\hsize}{!}{\includegraphics{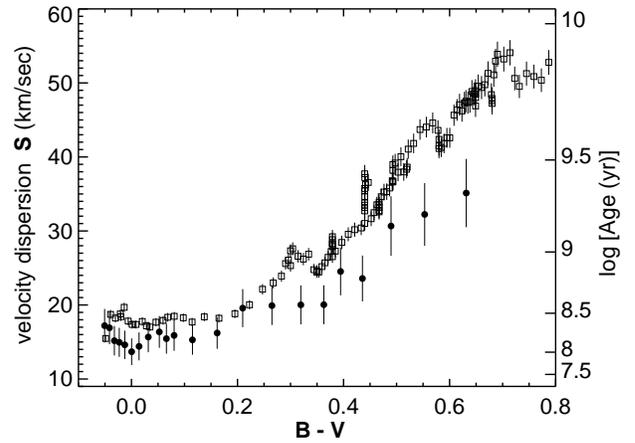}}
\caption{\sf Transverse velocity dispersion $S$ plotted against the
dereddened color $(B-V)$ for Vega-like stars and normal field
stars. Filled circles represent Vega-like stars and open squares the
normal field stars. The vertical error bars are errors in velocity
dispersion. The ages derived from the velocity dispersion using the
formalism of \citet{bdb00} is also shown on the y-axis on the
right.  Note that the age axis is not linear}
\label{fig1}
\end{figure}

In Fig~\ref{fig1} we plot the velocity dispersion of Vega-like stars
against their dereddened $(B-V)$ color. The values of transverse
velocity dispersion S shown in Fig~\ref{fig1} are for a sliding window
of 30 stars plotted against the mean $(B - V)$ for each bin. A fresh
point is plotted every time six stars have left the window. We have
also plotted the velocity dispersion of the field stars, computed from
their {\it Hipparcos} proper motions and parallaxes in the same way as
that for Vega-like stars, for comparison. The values of
S plotted for field stars are for bins of 500 stars with a new point
plotted when 100 stars have left the window. The error bars plotted in
S are the standard deviation of dispersion in each bin which is given
by $\Delta S\:=\:S/\sqrt{(2n-2)}$ where $n$ is the number of stars in
the bin. 

It can be seen from Fig~\ref{fig1} that the Vega-like stars, on 
average, show a lower velocity dispersion than the field stars at any
given (B - V). While this trend is clearly evident on visual
inspection for stars with $B - V\:\ge \:0.3$ (spectral type F0 or
later), it is not as striking for stars of early spectral type, though
in general, their velocity dispersions are smaller than the field
stars. A two-sided two dimensional Kolomogorov-Smirnov test shows that
the velocity dispersion - (B - V) relation for Vega-like stars and
field stars to be different with probability 99.99\%.  Thus, Vega-like
stars have a lower velocity dispersion than that of field stars for any
given spectral type. Since a lower velocity dispersion indicates younger
ages, it follows that the main sequence stars with debris disks are
statistically younger than the field stars of similar spectral
type. It can also be seen from Fig~\ref{fig1} that late type Vega-like
stars have a larger velocity dispersion than early type Vega-like stars,
suggesting that statistically they are older.

In order to quantify the disk lifetimes of Vega-like stars, we relate
the velocity dispersion to the stellar age following the formalism of
\citet{bdb00} where the velocity dispersion for a coeval group of
stars as a function of age is given by \( S\: ={v_{10}[(\tau +\tau
_{1})/(10Gyr+\tau _{1})]^{\beta }} \) . In this equation, \( \tau _{1}
\) determines the random velocity of stars at birth, and \( v_{10} \)
and \( \beta \) characterize the efficiency of stellar
acceleration. Using values of \( \beta =1/3 \), \( v_{10}\) = 58 km
s$^{-1}$ and \( \tau _{1}~=~0.03\)~Gyr \citep{bdb00} we translate
velocity dispersion into age. The ages thus derived from the velocity
dispersion are plotted in Fig~\ref{fig1} on the y-axis on the
right. The figure shows Vega-like stars to be systematically younger
than the field stars for all values of $(B - V)$.

Vega-like stars being a younger population than the field stars
indicates that the lifetimes of debris disks are shorter than the main
sequence lifetimes of the stars which harbour them.  The debris disk
may not survive for the entire lifetime of the central star. However,
it is possible that this is a selection effect, as {\it IRAS} will not
be able to detect low luminosity disks below its sensitivity
limit. The luminosity of debris disks gradually goes down as the
central stars age and eventually falls below the limit of
detection. This is expected physically as the larger bodies which
replenish the disk are eroded continuously and are finite in
supply. We also find such a fall in fractional dust luminosity of
debris disks with age as demonstrated in Fig~\ref{fig2} (see $\S$4.2).
Dust debris of low optical depth $f_d\:\la10^{-7}$ like that around
the Sun may be present over the entire lifetimes of the main sequence
stars.

 The ages of Vega-like stars are found to range from $10^8$ yr to
 $1-2$ Gyr with Vega-like stars of later spectral types being older on
 average than stars of early spectral types. Ages of early type
 Vega-like stars span from $10^8$ yr to $4-5\times 10^8$ yr while late
 type Vega-like stars can be as old as $1-2$ Gyr. That  main sequence dusty
 systems of earlier spectral type are systematically younger than
 the late type systems strongly suggests shorter lifetimes for debris
 disks around early type stars. This would mean that the temporal
 evolution of main sequence disks is a function of the spectral type
 of the central star, with debris disks surviving longer around late
 type stars, a conclusion also reached by \citet{songasp01}.  This
 is not surprising as the timescales for grain removal processes like
 radiation pressure ` blowout' and Poynting-Robertson drag which
 dominate Vega-like disks are inversely proportional to the stellar
 luminosity, and thus are shorter for early type stars.

\subsection{Temporal evolution of dust disks}

\begin{figure}
\centering
\resizebox{\hsize}{!}{\includegraphics{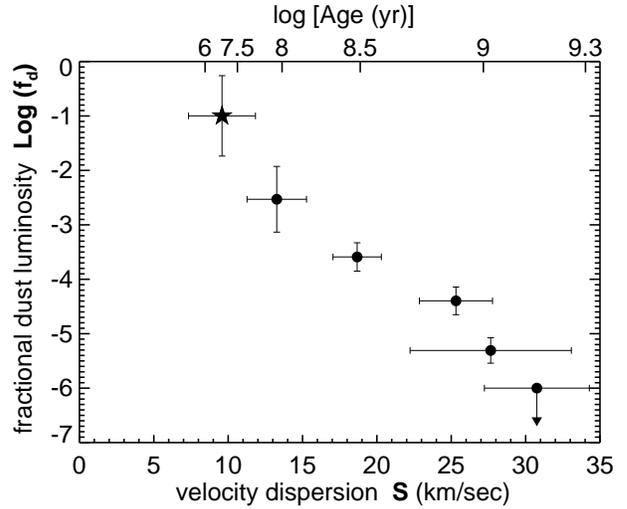}}
\caption{\sf Fractional dust luminosity $f_d$ plotted against
transverse velocity dispersion $S$ for stars with circumstellar dust
disks. Filled circles represent Vega-like stars. Herbig Ae/Be stars
are also plotted, represented by the solid star symbol.  The
point with a downward arrow represents stars with f$_d$ $\le$~
10$^{-6}$ (see text).  Horizontal error bars represent errors in
velocity dispersion.  The ages appropriate for the velocity dispersion
are also shown on the x-axis on the top.  Note that the age axis is
not linear }
\label{fig2}
\end{figure}

                  Next, we  study the evolution of the
`dustiness' of circumstellar disks with age. We look for a correlation
between fractional dust luminosity \( f_{d} \) and transverse velocity
dispersion S, for stars of similar \( f_{d} \). For this we grouped
the stars into bins of a given range in dustiness (\( f_{d} \)).  We
then computed the dispersion in transverse velocities of stars in each
of these bins. The mean value of \( f_{d} \) in each bin is then
plotted against the velocity dispersion of stars in that bin as shown
in Fig ~\ref{fig2}. Error bars for \( f_{d} \) represent the standard
deviation from the mean in each bin. Error bars in S are computed in
the same way as that in Fig ~\ref{fig1}.  The point with a
downward arrow represents all stars for which the computed f$_d$ $\le$
10$^{-6}$ (see $\S$3.1).  In Fig~\ref{fig2}, we also include Herbig
Ae/Be stars which are pre-main sequence stars of intermediate mass and
are thought to be the progenitors of Vega-like stars. We computed \(
V_{T} \) and \( f_{d} \) for 44 Herbig Ae/Be stars taken from
\citet{the94}, \citet{ van98} and \citet{malf98}, and for which
\textit{Hipparcos} and \textit{IRAS} measurements are available. As
for Vega-like stars, we restrict ourselves to stars within 250 pc and
with fractional errors in transverse velocities less than \( 0.5
\). We have 22 such Herbig Ae/Be stars.  Their average \( f_{d} \) is
plotted against the velocity dispersion S and is represented by the
filled star symbol in Fig~\ref{fig2}.

      It is clear from Fig~\ref{fig2} that there is a systematic
decrease in the dustiness (\( f_{d} \)) of the disks with increasing
velocity dispersion \( S \) of stars. Herbig Ae/Be stars have the  velocity
dispersion $S$ smaller than that of  Vega-like stars, indicating younger
ages. As discussed earlier, the velocity dispersion S of stars, in
general, is found to increase with stellar age as \( S\: \propto \:
t^{1/3} \) \citep{bdb00}.  The correlation between fractional dust
luminosity of stars with disks and their velocity dispersion seen in
Fig~\ref{fig2} clearly implies a steady decrease in the optical
thickness of the disks with stellar age. This is consistent with the
earlier findings that the amount of dust in the disks appears to
decrease generally with system age.

    The stellar ages obtained from translating velocity dispersion
into age using the formalism of \citet{bdb00} are also shown in
Fig~\ref{fig2} on the x-axis on the top.  The steady drop of
fractional dust luminosity with increasing stellar age is evident from
the figure. There is an overall decrease in the `dustiness' of the
circumstellar disks from the early pre-main sequence phase ( a few
Myr) to well up to the late main sequence phase ($\sim$ 1 - 2 $Gyr$).
It can also be seen from Fig~\ref{fig2} that the Herbig Ae/Be stars
have larger values of $f_d$ and younger ages than Vega-like
stars. This is consistent with them being the progenitors of Vega-like
stars.

\begin{figure}
\centering
\resizebox{\hsize}{!}{\includegraphics{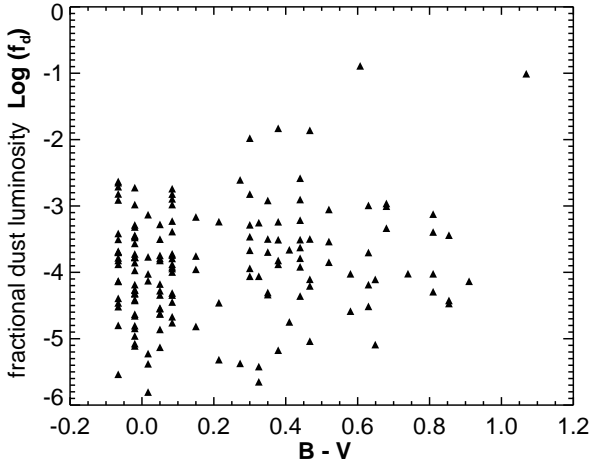}}
\caption{\sf Fractional dust luminosity $f_d$ plotted against
dereddened color $(B-V)$ for Vega-like stars }
\label{fig3}
\end{figure}

 We note here that  the correlation that we find between $f_d$ and
velocity dispersion is not because of the spectral type dependence of
velocity dispersion that we discussed earlier. Such a manifestation is
expected if $f_d$ has a spectral type dependence where the late type
stars preferentially have lower values of $f_d$. However, for our
sample stars, we find that there is no such trend of $f_d$ with
spectral type. In Fig~\ref{fig3} we present a plot of $f_d$ against
$(B-V)$ where it can be seen that there are as many or more early type
stars with lower values of $f_d$ as there are late type stars.

It should also be noted that the ages that we derive from the velocity
dispersion are statistical in nature. They are the average ages
appropriate for the velocity dispersion shown by the group of stars
and are derived from the relation between velocity dispersion and age
as given by \citet{bdb00}. The ages obtained this way can have
relatively large errors for small ages because for small ages the  velocity
dispersion $S$ is a steeply rising function of age for the relation
$S\:\propto\:t^{1/3}$.

\section{Discussion}

 Of the 158 stars in our final sample for which the analysis was
carried out, 107 (68\%) are within 100 pc and 51 (32\%) have d $\gt$
100 pc.  \citet{kalas02} have cautioned that the far-infrared emission
from some of the Vega-like stars beyond 100 pc could be thermal
emission from interstellar reflection nebulosities, similar to that
seen in the Pleiades, due to the chance encounters of stars with
relatively dense interstellar clouds. The dust density $\rho$
of the reflection nebulosities required to account for the observed fluxes is
in the range of 10$^{-24}$ - 10$^{-23}$ gcm$^{-3}$ \citep{kalas02}
which corresponds to a gas number density $n$ of 10$^2$ -10$^3$
cm$^{-3}$. Such high densities are generally associated with molecular
clouds and not with the normal interstellar HI clouds. In fact the
reflection nebulosity which causes the so called 'Pleiades phenomenon'
is a fragment of the Taurus-Auriga molecular cloud that has been
encountered by the Pleiades in that cluster's southward motion
\citep{hs01}. Such encounters of stars with dense molecular cloud
clumps in the solar neighbourhood must be rare. Of the initial sample
of 79 Vega-like candidates of \citet{kalas02}, 72 have {\it
Hipparcos}-detected distances, and of these 43 (60\%) have d $\gt$ 100
pc.  Among the 60 stars that they have observed from this sample, only
six are found to have reflection nebulosities surrounding them. We
have excluded these stars from our sample. Therefore amongst the stars
in our sample that have not yet been investigated for the `Pleiades
phenomenon', one would expect only a few additional objects with
reflection nebulosities. Because our analysis is statistical in
nature, the average trends that we derive are not expected to be
affected, even if a few stars with d $\gt$ 100 pc are found to be
surrounded by reflection nebulosities instead of orbiting dust
grains. Restricting our entire analysis to  stars within 100 pc gives
the same trends and does not affect the results, except  for poorer 
number statistics.

 In Fig~\ref{fig2} we have plotted both Herbig Ae/Be
 stars and Vega-like stars together and have derived an overall
 decline in fractional dust luminosity with stellar age. We point out
 here that the pre-main sequence disks like those around Herbig Ae/Be
 stars and the main sequence disks are physically different. Pre-main
 sequence disks are relatively optically thick ($f_d\sim0.1$) and
 gas-rich and are formed from the primordial cloud core from which the
 star itself is born. The infrared excess that the {\it IRAS} detected
 in these disks is due to the re-radiation from the first generation
 dust grains. On the other hand, the infrared excess shown by
 Vega-like disks is due to the debris dust produced in the collisions
 between larger bodies. The debris disks are gas-poor and the disk
 evolution is dominated by dust dynamics. Moreover, the decline in the
 fractional dust luminosity in the pre-main sequence disks is
 primarily due to the grain growth by which smaller sub-micron sized
 grains get depleted in the disks thereby reducing the effective
 surface area of dust absorption/emission. In main sequence debris
 disks, where the grains causing infrared excess are
 continuously being replenished, the fall in $f_d$ is due to the decline in
 the collisional regeneration rate. 

 Nevertheless, it is expected that the pre-main sequence disks
 gradually evolve into Vega-like disks, although it is not yet clear
 when exactly the secondary dust generation begins in these
 disks. Recent studies have suggested that in general, the lifetimes
 of primordial disks are only a few Myr
 \citep{haisch01,lada99}. Larger bodies like kilometer-sized
 planetesimals and comets, which replenish the main sequence debris
 disks, can also be formed within a few Myr
 \citep[e.g.][]{beck00}. Thus the transition of optically thick disks
 into optically thin disks is expected to take place on similar
 timescales. The resolution of the stellar ages derived from velocity
 dispersion is poor and is inadequate to address the issue of the
 timescale of the transition from primordial pre-main sequence disks
 to Vega-like disks. However, a general decline in the `dustiness' of
 the disks with stellar age from the pre-main sequence phase to the
 late main sequence phase is clearly seen.

  The results that we obtain from our analysis
are consistent with the earlier studies on the disk evolution around
main sequence stars. The overall decline in the dust content of the
debris disks has been reported by \citet{zb93}, \citet{hol98} and
\citet{spang01}. The conclusions of \citet{decin03} that there are few
young stars with small excesses and that a Vega-like excess is more
common in young stars than in old stars are consistent with our
result of Vega-like stars being younger than the field stars for all
spectral types. \citet{decin03} also argue that at most ages, there is
a spread in fractional dust luminosity of Vega-like stars. While this
may be true, our results strongly suggest that there is also a general
decline in the $f_d$ of the disks with stellar age.  However,
we are not able to fit a single power law to the fall in $f_d$ with
the average stellar ages that we derive from the velocity dispersion of
the stars.

\section{Conclusions}

In this paper we have used  velocity dispersion as an age indicator
to constrain the ages of a large sample of Vega-like stars.  From the
statistical ages derived from the velocity dispersion, we have studied
the disk lifetimes and the temporal evolution of the dust disks around
main sequence stars.  The conclusions of this study are summarized
below.

\begin{itemize}
\item{} Velocity dispersion of Vega-like stars is found to be smaller
than that of main sequence field stars for all spectral
types. Main sequence stars with debris disks, on average, are
younger than  normal field stars of similar spectral type.

\item{} The ages of Vega-like stars derived from the velocity dispersion
 range from $10^8$ yr to $1-2$ Gyr.

\item{} Vega-like stars of later spectral types are statistically
older than Vega-like stars of earlier spectral type. Debris disks seem
to survive longer around late type stars as compared to early type
stars.

\item{} There is a strong correlation between fractional dust
 luminosity and velocity dispersion of Vega-like stars. Average
 fractional dust luminosity $f_d$ of stars with disks decreases
 monotonically with increasing velocity dispersion.  There is a
 general decline of $f_d$ with stellar age  from the  early pre-main
 sequence phase to the late main sequence phase.

\item{} The observed high $f_d$, lower velocity dispersion $S$ and the
implied younger ages for Herbig Ae/Be stars are consistent with them
being the progenitors of Vega-like stars.

\end{itemize}

\bibliography{references}
\bibliographystyle{aa}

\end{document}